# In-plane Density Gradation of Shoe Midsoles for Optimized Cushioning Performance

Kazi Zahir Uddin[a,*], Hai Anh Nguyen[a,b], Thanh T. Nguyen[b], Mitja Trkov[a],
George Youssef[c], Behrad Koohbor [a,d,*]

a. Department of Mechanical Engineering, Rowan University, 201 Mullica Hill Rd., Glassboro, NJ 08028, USA
b. Department of Mathematics, Rowan University, 201 Mullica Hill Rd., Glassboro, NJ 08028, USA
c. Experimental Mechanics Laboratory, Department of Mechanical Engineering, San Diego State University, 5500 Campanile Drive, San Diego, CA 92182, USA
d. Advanced Materials and Manufacturing Institute, Rowan University, 201 Mullica Hill Rd., Glassboro, NJ 08028, USA
* Corresponding Authors, Emails: uddink79@students.rowan.edu & koohbor@rowan.edu   |   Phone: +1 (856) 256-5328

**Abstract**

Midsoles are important components in footwear as they provide shock absorption and stability, thereby improving comfort and effectively preventing certain foot and ankle injuries. A rationally tailored midsole can potentially mitigate plantar pressure, improving performance and comfort levels. Despite the importance of midsole design, the potential of using in-plane density gradation in midsole has been rarely explored in earlier studies. The present work investigates the effectiveness of in-plane density gradation in shoe midsoles using a new class of polyurea foams as the material candidate. Their excellent cushioning properties justify the use of polyurea foams. Different polyurea foam densities, ranging from 95 to 350 kg/m$^3$ are examined and tested to construct density-dependent correlative mathematical relations required for the optimization process. An optimization framework is then created to allocate foam densities at certain plantar zones based on the required cushioning performance constrained by the local pressures. The interior-point algorithm was used to solve the constrained optimization problem. The optimization algorithm introduces a novel approach, utilizing the maximum specific energy absorption as the objective function. The optimization process identifies specific foam densities at various plantar regions for maximum biomechanical energy dissipation without incurring additional weight penalties. Our results suggest that midsole design can benefit from horizontal (in-plane) density gradation, leading to potential weight reduction and localized cushioning improvements. With local plantar peak pressure data analysis, the optimization results indicate low-density polyurea foams (140 kg/m$^3$) for central and lateral phalanges, whereas stiffer foams (185-230 kg/m$^3$) are identified as suitable candidates for metatarsal and arch regions in an in-plane density graded midsole design. The approach presented herein has the potential to be applied to a wide range of gait speeds (loading rates) and user-specific plantar pressure patterns for enhanced functionality and cushioning performance.

## 1. Introduction

Foot-related disorders and diseases, including osteoarthritis, flat feet deformity, diabetes, and plantar ulcers, are prevalent and can lead to pain, disability, and reduced quality of life.[1,2] These conditions are widespread among the elderly and can cause imbalance and discomfort, resulting in additional risks associated with concomitant bone and muscular diseases.[3-5] Although clinical treatments such as medication or surgery can be effective, they are often costly and increase postoperative risks.[6,7] Alternatively, non-pharmacological interventions like custom-made orthopedic shoes can offer practical approaches to mitigate plantar pressure, improve functionality and balance, and reducing pain.[8,9] Beyond therapeutic application, a customized footwear design can improve athletic performance and minimize sports related injuries. Appropriately designed shoes can alter the stress distribution on the foot during exercise, potentially leading to improved biomechanics and injury prevention.

The cushioning response of footwear is crucial for both comfort and optimal biomechanics characteristics. Proper cushioning improves mechanical energy dissipation and plays an essential role in redistributing applied forces on the foot, effectively reducing the risk of lower limb injuries.[10, 11] Shoe sole stiffness, a critical metric for user comfort, can be altered through custom-made insoles to mitigate high plantar pressures.[12, 13] Despite the benefits of custom insoles for pressure attenuation and stress distribution, their relatively thin structure limits their effectiveness, particularly during high-impact activities such as jogging or running.[14, 15] In such conditions, the shoe midsole predominantly dissipates most of the biomechanical impact forces due to its larger thickness than other components in footwear.[16] While the present work focuses on optimizing midsole density, other footwear components also play critical roles in improving comfort and functionality. In fact, specific insole designs, despite being relatively thinner than midsoles, have been shown to be effective in mitigating ground impact forces during physical activities. For instance, a study by Chen et al. demonstrated the effectiveness of insole thickness and metatarsal pad placements in reducing plantar pressure using finite element analysis.[17] Another study showcased the potential of a conforming heel cup and softer insole for enhanced running shoe cushioning with plantar pressure relief.[18] Meanwhile, the incorporation of carbon plates in midsole structural design has shown a significant influence on impact force distribution and lower limb biomechanics during exercise. Running shoes with carbon-fiber plates can alter planter pressure, and affect running biomechanics, thereby having implications on running efficiency and injury risk.[19] From an engineering perspective, the microstructure of the midsole material predominantly contribute to its load-bearing and mechanical energy absorption characteristics.[20] As a result, effective shoe midsole design necessitates a multidisciplinary approach, considering the symbiotic interrelationships between the material properties, mechanical performance, and the biomechanical conditions of footwear.

In line with the recent advancements in additive manufacturing, researchers have explored the potential of lattice structures in shoe midsole design,[21-24] benefiting from their exceptional load-absorbing and lightweight nature.[25,26] By modifying the unit cell topology and architecture, lattice structures such as honeycombs, triply periodic minimal surfaces (TPMS), and Voronoi-inspired geometries can achieve adjustable deformation behaviors and tunable compressive properties.[27-29] This includes the innovative 'auxetic metamaterials',[30-33] already implemented in commercially available Nike-Auxetic running shoes.[34] However, customizing lattice structures for specific midsole users requires a meticulous design strategy due to biomechanical sensitivity, typically achieved through various optimization techniques.[35-39] The design optimization algorithms utilized thus far are plagued with several drawbacks, including a short lifecycle caused by strain localization at intersecting cell walls in a lattice structure. This leads to poor cushioning performance over long-term use.[40,41] Despite the advancements in lattice-designed midsoles, their limitations have led researchers to explore alternative midsole customization and optimization approaches. Polymeric foam designs, particularly modular and multi-density structures with structural hierarchy, have emerged as promising alternatives for next-generation footwear, especially in athletic applications.[42,43] In particular, the emergent design concepts revolving around the functional gradation of energy-absorbing structures are gaining attention in the footwear industry.

Functionally graded materials (FGMs), in general, provide superior performance due to their tailored spatial distribution of key performance indicators (including mass density), satisfying local mechanical behavior requirements.[44-46] FGMs are exemplified by a gradual variation in composition, microstructure, or properties, providing improved performance and adaptability.[47,48] For example, Shimazaki et al. demonstrated that functionally graded ethyl vinyl acetate (EVA) foams outperformed uniform-density counterparts in absorbing higher impact energy exerted on midsoles during normal walking.[20] Several additional studies also documented enhanced energy absorption in functionally graded foam materials (FGFM) compared to their monolithic counterparts,[44,49] including their effectiveness in further reducing peak plantar pressure by increasing the foot-shoe sole contact area.[50] Consequently, functionally graded foam structures have been shown to provide higher durability, cushioning performance, and impact energy absorption than lattice-based structures, *i.e.*, FGFMs are an ideal choice for improved footwear comfort and cushioning. Furthermore, density gradation within the midsole structure is a practical strategy to reduce the overall weight of the shoes and achieve tailorable cushioning and mechanical performances while improving running economy (*e.g.*, reducing oxygen consumption).[51,52] For example, it has been demonstrated that horizontally graded (segmented) shoe midsoles effectively redistribute plantar foot pressure, improving forefoot stability and comfort.[53,54] Additionally, research conducted by the current authors indicates that density-graded polyurea foams surpass single-density counterparts in terms of combined mechanical load-bearing and energy absorption capacities, offering biomedical advantages for the design of new orthopedic shoes.[48,55] This concept can be extended to achieve user-specific, non-conventional orthotics and athletic footwear by personalizing functionally graded midsole density and stiffness.

The current research addresses the challenge of designing horizontally (in-plane) graded shoe midsoles for optimal footwear performance by leveraging the promising potential of density-graded foam structures. The systematic approach developed here hinges primarily on advancing the understanding of density-graded foam structures in footwear design emphasizing personalization. By developing property-adjustable polyurea foams and utilizing a mathematical optimization algorithm, we seek to bridge the gap between the mechanical properties of foam and biomechanics in order to identify local mechanical properties that result in truly optimal footwear designs. In addition, we aim to establish a solid foundation for validation and comparison with the novel polyurea foam structures explored in this study by conducting experimental measurements on commercially available shoe midsoles to understand their mechanical deformation.

## 2. Experimental Protocol

### 2.1. Materials and Mechanical Testing

Flexible EVA and polyurethane foams are the most widely used midsole material in walking and running shoes.[56-58] Polyurea foam has recently been proposed as a potential material candidate for footwear due to its excellent cushioning, water resistance, ease of manufacturing, and tailorable properties.[59,60] In addition, recent studies suggest that polyurea foams outperform others in impact mitigation by an extended plateau region in their stress-strain response.[59,61,62] This unique mechanical behavior enhances energy absorption and resilience, making polyurea foams particularly suitable for applications that require high mechanical energy absorption, such as footwear midsoles. Therefore, this work uses polyurea foams with different nominal densities to design and optimize graded midsoles.

The examined polyurea foams (hereafter referred to as EML foams) were fabricated based on the manufacturing process reported by Reed et al.[59] in three relative densities of 0.095, 0.23, and 0.35. Complete details regarding foam fabrication can be found in.[59,62,63] The term 'relative density' is defined as the ratio of the actual density of the foam to the density of water at 20°C.[55] The mechanical and energy absorption performances of the three EML foams (referred to as EML 95, EML 230, and EML 350 herein) were characterized by quasi-static compression. The strain rate conditions used to test the foam samples in this work resembled the loading conditions present in normal walking. These strain rate conditions were validated by performing *in situ* digital image correlation (DIC) measurements on actual midsoles, as discussed in **Sec. 4.1**. For comparison purposes, stress-strain and energy absorption responses of a commercial midsole foam material were also measured by conducting similar compression tests. Cubic foam samples with dimensions 1×1×1 cm$^3$ were extracted from the midsole of a Nike Revolution 3® walking shoe. The material used in the midsole of this shoe is reported to be a phylon (compressed and heat-expanded EVA) foam, with a nominal density of 177 kg/m$^3$ (measured in-house).

The specific energy absorption capacity of the foams, *W*, was determined by calculating the area under the stress-strain curve as

$$W(\varepsilon) = \int_0^\varepsilon \sigma(\varepsilon) \cdot d\varepsilon, \tag{1}$$

where, $\sigma$ and $\varepsilon$ denote compressive engineering stress and strain, respectively.[55] The *ideality* metric was then used to compare the mechanical energy absorption performance of the utilized foam samples to that of an 'ideal energy absorber.' The ideality metric, *I*, is defined as the ratio between the energy absorption and the product of stress and strain,[55] expressed as

$$I(\varepsilon) = \frac{\int_0^\epsilon \sigma(\varepsilon) \cdot d\varepsilon}{\sigma \cdot \varepsilon} \tag{2}$$

An ideal absorber is one for which *I* is unity.

**2.2. *In Situ* Measurement of Strain and Strain Rates in Midsoles**

Benchmark experiments were designed and conducted to determine the nominal strains and strain rates applied on shoe midsoles during normal walking conditions. Midsole testing informed the design of experiments to determine the stress-strain and energy absorption response of EML and benchmark EVA foams. The strain-related attributes were quantified by measuring the strain fields developed on the midsole of a Nike Revolution 3® during normal walking. The outer surface of the midsole was coated with a black-and-white speckle pattern to facilitate DIC.[64] The subject (85 kg, male) was asked to walk normally while a high-speed camera (Grasshopper GS3-U3-51s%m-C, FLIR, OR, USA; Imaging rate: 30 fps) recorded the time-lapse images of the midsole during one complete stance phase of a gait cycle. Strain fields developed on the exterior of the midsole side, underneath the heel area, were resolved using commercial DIC software Vic-2D (Correlated Solutions, Inc. SC, USA). The spatial average of the axial compressive strains within the area of interest was plotted as a function of time. The slope of the best linear fit to the loading half-cycle was taken as the nominal strain rate, which was then used to design the mechanical tests performed on the foam samples, as discussed in **Sec. 3.1**.

### 2.3. Plantar Pressure Measurements

Local compressive stresses applied to the midsole were measured using pressure-sensitive insoles (Moticon, Germany) during normal walking (~5 km/h). The subjects used for plantar pressure measurements (79 kg, male) and midsole strain measurements were different individuals due to the differences in the time and location of the two measurements. The insole sensors collected local stress histories at various plantar locations. The range of measurement pressures was within 0-0.4 MPa at a resolution of 2.5 kPa. Plantar pressure data was collected at a rate of 50 Hz. The local pressure peaks were extracted and used as input to the optimization model discussed in the following sections. More details regarding plantar pressure measurement protocols can be found in.[55]

## 3. Model Development

### 3.1. Density-Dependent Constitutive Response of Polyurea Foams

Polyurea foams EML 95, EML 230, and EML 350 were fabricated and subjected to mechanical testing at quasi-static loading rates. The experimental stress-strain curves were fitted into $5^{th}$-order polynomials with the general form,

$$\sigma(\varepsilon) = a\varepsilon^5 + b\varepsilon^4 + c\varepsilon^3 + d\varepsilon^2 + e\varepsilon \tag{3}$$

where, $a, b, c, d$, and $e$ are density-dependent coefficients. While a fifth-order polynomial, **Eq. 3**, has been used in this work to demonstrate the viability and applicability of the enclosed algorithm, the approach is rather inclusive of any mechanistic or mathematical representation of the stress-strain curve. That is to say; the same approach can readily be applied irrespective of the resulting mechanical behavior by using any hyperelastic constitutive model.[65] Once the fitting parameters were extracted after the regression analysis, they were individually related to the density through a secondary regression of the fitting parameters (*e.g.*, $a, b, c, d,$ and $e$) as functions of density using a quadratic polynomial for simplicity. **Table 1** shows the empirical mathematical formulae for density-dependent coefficients (*a*, *b*, *c*, *d*, and *e*) in **Eq. 3**. At this point, the stress-strain response of any polyurea foam with an arbitrary density between the upper and lower bounds can be easily predicted through interpolation.

**Table 1-** Mathematical expressions of the density-dependent coefficients used to describe the stress-strain response of polyurea foams with various densities. The units of $\rho$ in all equations is kg/m$^3$.

| Coefficient | Expression |
|---|---|
| a | $3.64 \times 10^{-4}\rho^2 + 0.065\rho + 11.91$ |
| b | $1.66 \times 10^{-5}\rho^2 - 0.24\rho - 11.25$ |
| c | $-1.79 \times 10^{-4}\rho^2 + 0.17\rho + 5.24$ |
| d | $8.6 \times 10^{-5}\rho^2 - 0.052\rho - 0.798$ |
| e | $-6.92 \times 10^{-6}\rho^2 + 0.006\rho + 0.08$ |

**Figure 1a** shows the model-predicted mechanical response and experimentally measured stress-strain curves from the three nominal-density foam samples. Based on the identified fitting model parameters, density-dependent constitutive maps were constructed. **Figure 1b** shows the variation of density-dependent stress-strain responses for physical and hypothetical polyurea foams. This figure substantiates the previously reported mechanical behavior of polyurea foams, *i.e.*, extended plateau region and enhanced energy absorption performance. Specifically, **Figure 1b** reveals that the induced stresses are generally lower than 1 MPa, for a broad range of strains and foam densities, which is attributed to the hyperelastic response of polyurea foams.[55] Furthermore, the extended plateau region, irrespective of the density, is axiomatic from the dominance of low stresses over nearly the entire strain range. The stress exceeds 1 MPa under only two conditions: (1) relatively high density and (2) under high strains, *i.e.*, nearly the end of the compression event.

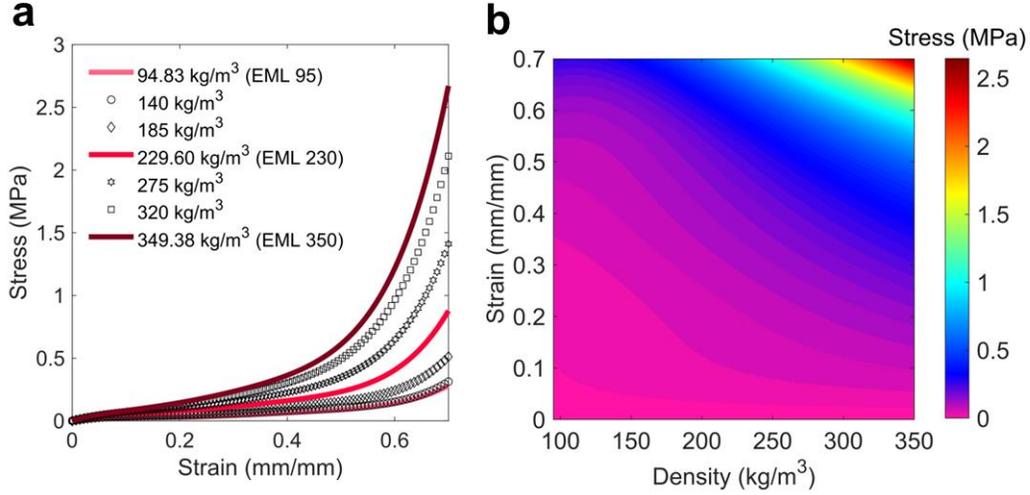

**Figure 1**. (a) Comparison of the stress-strain curves obtained experimentally (represented by solid lines) and predicted numerically (represented by symbols). (b) Contour map showing the variation of the density-dependent stress-strain response of polyurea foams, used as the input to the optimization algorithm. All stress-strain curves were obtained from quasi-static compression at a nominal strain rate of 0.06 s$^{-1}$.

## 3.2 Analytical Modeling and Optimization Approach

This section presents an analytical optimization approach aimed at determining the optimal structure or gradation sequence of density-graded foams with vertical (through-thickness) and horizontal (in-plane) gradations that maximize the ideality metric (**Eq. 2**) while accounting for the nonlinear stress-strain behavior of the polyurea foams (*e.g.*, **Figure 1a**). The basic assumption is that the shoe midsole consists of $N$ parallel layers of foams with known densities. As discussed in a previous study,[55] the assumption of a one-dimensional stress state remains valid, thus neglecting the presence of any shear strain/stress at the interfaces.[47] This assumption has been validated for the near zero Poisson's ratio of the base foams, discussed in detail elsewhere.[66] We first consider a uniaxial quasi-static compressive load applied on a foam laminate consisting of vertically stacked density layers (see **Figure 2**). In such a case, the applied global stress on the topmost surface remains spatially constant along the z-direction (*i.e.*, throughout the thickness).[44,50] The local density along the z-direction is set, while the local (layer-wise) strain associated with the compressive stress is unknown and differs between layers. In the case of nonlinear materials, such as the utilized polyurea foams, there is no direct method to determine the local strain value for a given stress-density pair. As such, no analytical correlation can be established between the performance attributes (*e.g.*, the ideality metric, **Eq. 2**) and the gradation sequence.

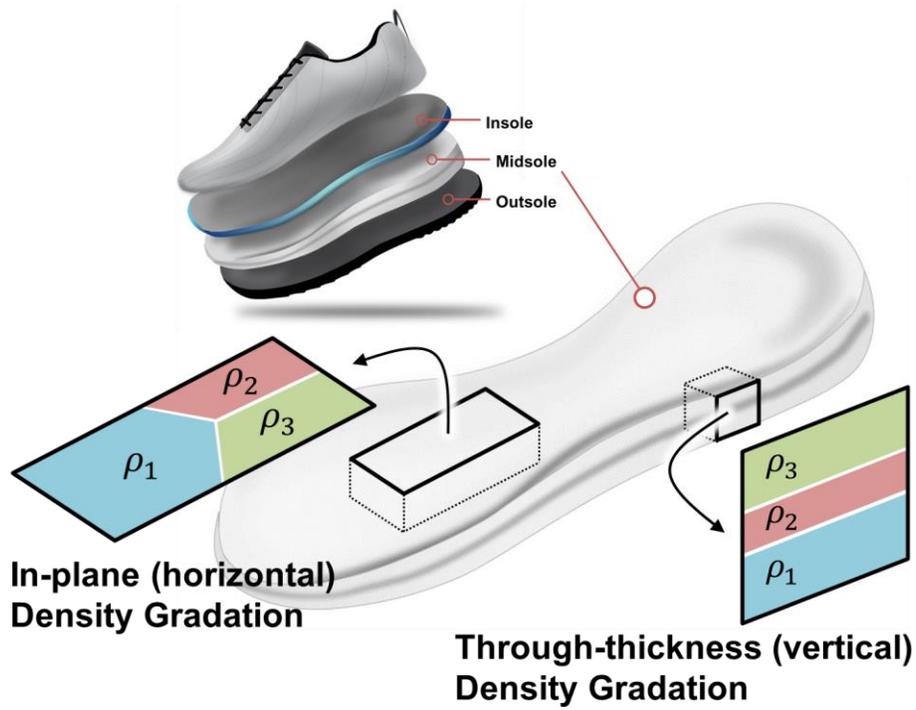

**Figure 2-** Schematic representation of the graded foam structures with vertical and horizontal gradients.

A mathematical optimization approach is proposed to address the challenges associated with the nonlinear foam response. Here, the thickness and density of the $n^{\text{th}}$ layer are denoted by $h_n$ and $\rho_n$, respectively. The layer-wise strain generated by any stress $\sigma$ is denoted by $\varepsilon_n(\sigma)$, which can easily be derived by interpolating an experimental stress-strain curve for that foam (see **Figure**

**1a**) using the cubic spline interpolation algorithm. Therefore, the strain caused by stress $\sigma$ on the $N$-layered structure is given by:

$$\varepsilon(\sigma) = ln\left(\frac{H}{\sum_{n=1}^{N} h_n e^{-\varepsilon_n(\sigma)}}\right) \quad (4)$$

It is physically reasonable to assume that $\varepsilon_n(\sigma)$ is a strictly increasing function, as shown earlier in **Figure 1a**, which reveals that each value of stress uniquely corresponds to one strain, *i.e.*, uniqueness is automatically granted since the experimental stress-strain curves are smooth without signs of elastic or inelastic collapse. Consequently, the strain $\varepsilon(\sigma)$ of the $N$-layered structure given by **Eq. 4** will also be a strictly increasing function of stress $\sigma$ for fixed thicknesses $h_1, h_2, \ldots, h_N$. Therefore, an inverse function $\sigma(\varepsilon)$ must exist. $\sigma_{max}$ denotes the maximum stress applied to the structure. To find the optimal structure associated with this stress, we assume that the layer densities are known, leading to amended formulae of the ideality of an $N$-layered structure,

$$I(h_1, h_2, \ldots, h_N) = \frac{\int_0^{\varepsilon_{max}} \sigma(\varepsilon) d\varepsilon}{\sigma_{max}\varepsilon_{max}} = 1 - \frac{\int_0^{\sigma_{max}} \varepsilon(\sigma) d\sigma}{\sigma_{max}\varepsilon_{max}} \quad (5)$$

where, $\varepsilon_{max}$ is the strain caused by the maximum stress $\sigma_{max}$. The last equality utilizes the inverse relations $\varepsilon(\sigma)$ and $\sigma(\varepsilon)$. The unknown parameters $h_1, h_2, \ldots, h_N$ are then found by maximizing the amended ideality function $I(h_1, h_2, \ldots, h_N)$ defined by **Eq. 5**. To do so, the integral in **Eq. 5** is approximated by a discrete sum applied by subdividing the interval $[0, \sigma_{max}]$ into $M$ subintervals by the grid points $0 = \sigma_0 < \sigma_1 < \cdots < \sigma_M = \sigma_{max}$. The strain is approximated by a piece-wise constant function with values on the subintervals being $\varepsilon(\sigma_0), \varepsilon(\sigma_1), \ldots, \varepsilon(\sigma_M)$. The ideality function in **Eq. 5** is approximated by the following discrete function, which is based on the trapezoidal quadrature rule.

$$I_d(h_1, h_2, \ldots, h_N) = 1 - \frac{\frac{1}{2}\sum_{n=1}^{N} [\varepsilon(\sigma_n) + \varepsilon(\sigma_{n-1})](\sigma_n - \sigma_{n-1})}{\sigma_{max}\varepsilon_{max}} \quad (6)$$

To find the thicknesses of the layers, one must consider the following constraints when the ideality function in **Eq. 6** is maximized.

$$h_1 + h_2 + \cdots + h_N = H \tag{7a}$$

$$h_n \geq h_{min} \tag{7b}$$

where, $h_{min} \geq 0$ represents the minimum thickness of each layer in the structure and $H$ is the total thickness of the structure. For any specific plantar location, the minimum layer thickness ($h_{min}$) is set to zero to minimize the foam stacking by removing unnecessary layers. This approach ensures that only the necessary vertical layer thicknesses for each foot location are obtained. As a result, the optimization approach maintains the density gradation concept while minimizing the number of vertical layers for improved structural integrity and performance. To reduce the number of unknown variables in the optimization problem, $h_N$ in **Eq. 7a** is represented through the other parameters as $h_N = H - h_1 - \cdots - h_{N-1}$. Hence, the constraint in **Eq. 7** is rewritten as:

$$h_1 + h_2 + \cdots + h_{N-1} \leq H \tag{8}$$

From **Eq. 6**, it follows that maximizing $I_d(h_1, h_2, \ldots, h_N)$ is equivalent to minimizing the function,

$$F(h_1, h_2, \ldots, h_{N-1}) = \frac{\frac{1}{2}\sum_{n=1}^{N}[\varepsilon(\sigma_n) + \varepsilon(\sigma_{n-1})](\sigma_n - \sigma_{n-1})}{\sigma_{max}\varepsilon_{max}}, \tag{9}$$

subject to the constraints (7b) and (8). This minimization problem is solved using the function *fmincon* in the MATLAB Optimization Toolbox to find solutions of constrained optimization problems.

The goal of the optimization process herein is to maximize energy absorption (*i.e.*, ideality) while minimizing the midsole weight. Two prominent energy dissipation and energy absorption efficiency (before the onset of densification) tend to increase with density, as reported extensively in our previous studies.[55] In contrast, the ideality does not monotonically increase with density, making it a more suitable parameter for balancing energy absorption and weight in the optimization process. This rationale further justifies the choice of ideality as the primary criterion for the optimization process in this work.

## 4. Results and Discussion

### 4.1. Benchmark Tests Performed on Commercially Available Footwear

**Figure 3a** shows the speckle-patterned area on the midsole and the region of interest (ROI) for strain measurements on a commercially available shoe during normal walking conditions. **Figure 3b** shows the evolution of the compressive strain field ($\varepsilon_{yy}$) within the ROI at various instances during a single gait. As shown in this figure, the top regions of the midsole undergo relatively larger compressive deformations, reaching local strain as high as ~0.1 during the loading phase of the gait. The highly heterogeneous strain distribution in the midsole challenges identifying a single strain value to represent the biomechanical deformation event. Therefore, the spatial averaging of the gait-induced mechanical strains can be obtained and considered for practical purposes. **Figure 3c** shows the evolution of averaged compressive strains over the ROI, showing the average strain monotonically increasing during the loading phase, *i.e.*, during the first ~ 500 ms. A full recovery is observed during the heel-off, unloading phase. Accordingly, the slope of the strain-time curve during the loading phase was determined to be 0.06 s$^{-1}$.

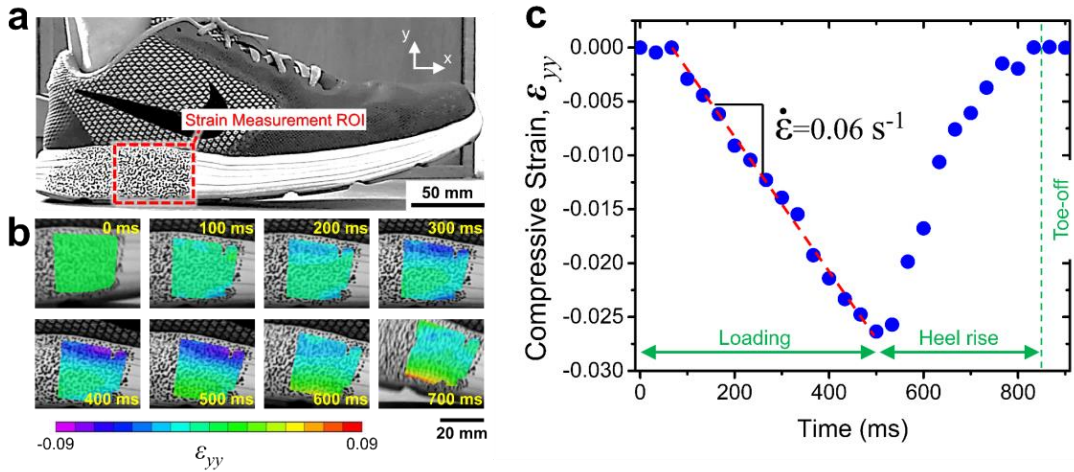

**Figure 3**. Development of biomechanical strains developed during normal walking in a Nike Revolution 3 shoe midsole, measured by digital image correlation: (a) Strain measurements in the region of interest (ROI), (b) compressive strain field developed near the heel area, and (c) temporal evolution of averaged compressive strain over the entire area of interest to resolve the strain rates.

In addition to the experimental measurement of shoe sole strain and strain rate, the performance attributes of the commercially available EVA foams were evaluated and compared with those of the polyurea foams considered herein. **Figure 4** compares the stress-strain and energy absorption ideality of the EVA foam with polyurea foams with three different densities. The stress-strain curves in **Figure 4a** indicate that the benchmark EVA foam possesses higher strength compared to the low densities of polyurea foams and delayed onset of densification strain. However, as shown in **Figure 4b**, the energy absorption ideality of the EVA foam is inferior to that of all polyurea foams. The maximum ideality of the EVA foam is shown to be achieved in compressive strains as high as 0.3, *i.e.*, strain values exceeding those measured during normal walking conditions (**Figure 3**). The ideality values for polyurea foams are significantly higher than the benchmark EVA foam, occurring at lower strains comparable to those generated during the walking event. This justifies the consideration of polyurea foams as a potential substitute for the incumbent footwear foams. The promise of polyurea foams has been further highlighted in other studies by excellent resistance to environmental conditions of the base materials, such as tolerance to extended ultraviolet radiation and accommodation of the foams to repeated impact conditions.[67-69]

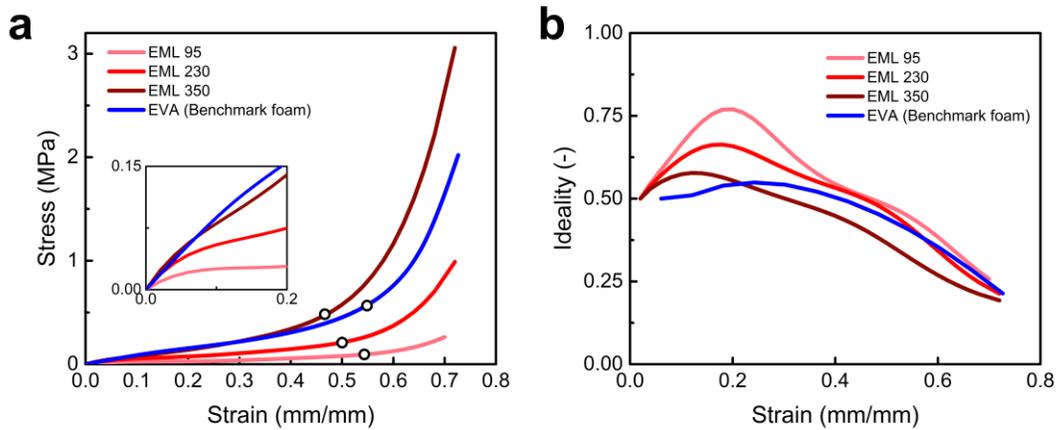

**Figure 4**. Comparison between (a) stress-strain and (b) ideality-strain behaviors of EML (polyurea) and benchmark EVA foams. The hollow circles in (a) mark the onset of densification strain for each material.

## 4.2. Optimal Gradation for the Graded Midsoles Design

The results in **Figure 4** indicate that the polyurea foams exhibit maximum ideality within a strain range of 0.15 to 0.2. In designing the optimized midsole structure, three base foam samples EML 95, EML 230, EML 350 and four virtual foam samples with densities of 140, 185, 275, and 320 kg/m$^3$ were employed. The stress-strain curves of the virtual foam samples were derived by interpolating the responses of the base foam samples (see **Eq. 3**). The selection of virtual densities was primarily based on equal increments in density between the base samples, providing a representative range of foam densities for the optimization process.

For the optimization process, all foam samples were considered as potential material candidates to design virtual foam laminates within a confined thickness. As midsole thickness varies across the foot, with a higher thickness in the rearfoot area compared to the midfoot and forefoot regions, we considered two midsole thicknesses: 30 mm and 50 mm. The input stresses were retrieved from the peak plantar pressure distribution on the midsole, as discussed in **Sec. 2.3**. **Figure 5** illustrates the optimized foam density corresponding to the stress distribution identified by the optimization process, where the latter opted for a single-layer midsole even though the multi-layer design was available. The optimized foam densities were also agnostic to the overall thickness of the midsoles, *i.e.*, 30 and 50 mm thick midsoles. That is, the optimization results indicate that a single-density foam is best suited for a specific plantar zone to accommodate the maximum applied pressure during walking. For example, the EML 95 foam material acts as the optimized choice for stresses up to 0.039 MPa among the seven foam densities. Similarly, for a stress range from 0.04 MPa to 0.047 MPa, the 140 kg/m$^3$ foam density is more suitable based on the ideality metric discussed above. Consequently, this finding indicates that the optimized midsole design can be characterized by a horizontal density gradation (in-plane distribution) instead of out-of-the-plane gradations.

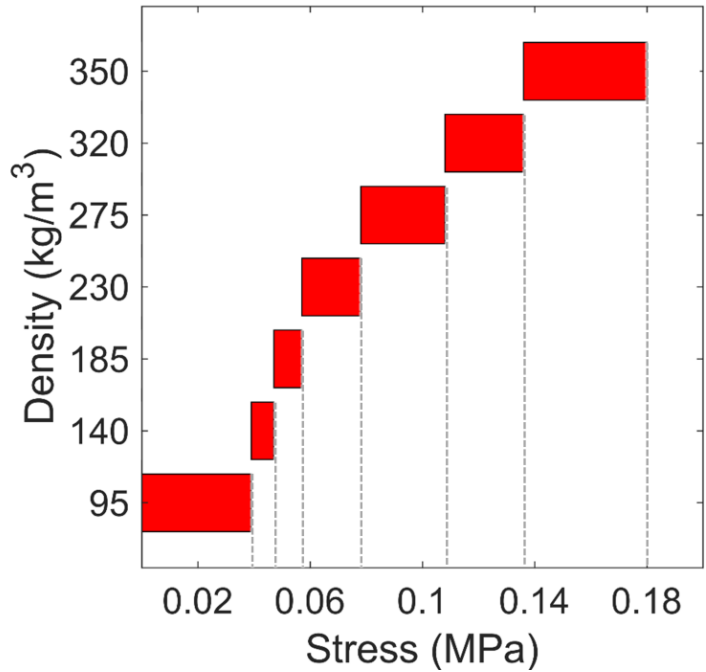

**Figure 5.** Optimized foam densities for various stress ranges. Note that the densities 95, 230, and 350 referred to the EML 95, EML 230, and EML 350 foams respectively. The graph shows a nonlinear correlation between the optimal foam density and the applied stress.

In-plane density gradation for shoe soles using polyurea foams provides two opportunities for improving existing orthotics and pursuing personalized solutions. First is the potential of substantially reducing weight by eliminating the need for insoles and outsoles, given the desirable mechanical and physical attributes of polyurea foams. A byproduct of the weight reduction benefit is easing the complexity of the manufacturing process. The second is delivering localized optimal cushioning and comfort performance of the soles based on the distribution of plantar pressures from different foot geometries and deformities. Moreover, the overall thickness of specific regions can be further optimized and minimized to accommodate biomechanical requirements such as maintaining proper joint kinematics during walking.

**4.3 Customizing Midsole Design Based on Plantar Pressure Distribution**

Developing effective and adaptive midsoles necessitates incorporating the local plantar peak pressure data, and identifying specific regions of the foot that experience maximum pressure during normal walking activities (*e.g.*, the hallux, medial, and central forefoot areas). To realize an in-plane optimal gradation, it was imperative first to identify the local plantar pressures, as illustrated in **Figure 6**. The midsole was divided into five separate areas based on the distinctive stress distribution patterns observed in the plantar pressure data. The optimal foam density identified for each pressure zone can be assigned to single or multiple regions. As such, the entire midsole is structured by systematically assigning the foam densities to their respective areas. The idea to not divide the midsole exactly as the high-pressure areas is justified by a balance between precision and practicality in the design process. This allocation process guarantees that the midsole design accommodates the unique biomechanics and pressure distribution patterns of each user, ultimately leading to enhanced footwear performance and comfort levels, measured herein by the ideality metric.

As visualized in **Figure 6**, the optimization process sought the lowest density foams with high ideality for the areas associated with the central/lateral phalanges (denoted by 'A' in **Figure 6**), enduring lower compressive pressures (see **Figure 5**). On the other hand, the entire heel area (marked as 'E' in **Figure 6**) and the regions associated with the medial phalanges and corresponding metatarsal (collectively denoted as segment 'B') require higher density foams to ensure sufficient mechanical energy absorption at higher pressure values. Finally, the optimal foam densities identified for the arch region (segments 'C' and 'D') are those with medium densities in the range of 185-230 kg/m³. This division is based on an average pressure distribution pattern observed across multiple subjects. Individual variations in foot pressure distribution can be quite significant. However, this design attempts to accommodate a broad range of users. The current study offers a preliminary insight into the feasibility of the approach and serves as a foundation for future studies that will consider a larger and more diverse subject population.

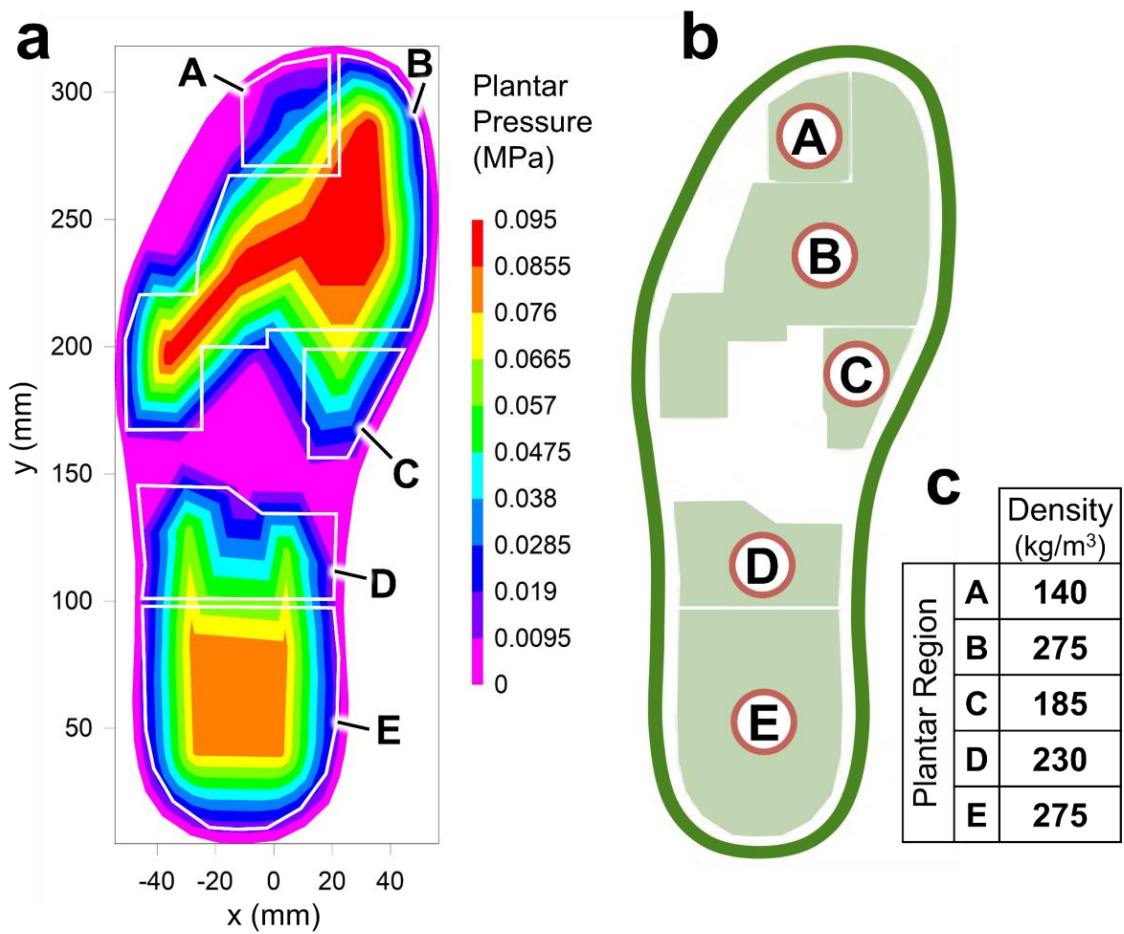

**Figure 6.** (a) Spatial distribution of the maximum plantar pressure for a 79 kg male subject during normal walking conditions. (b) In-plane optimal gradation of midsole foam densities based on plantar pressure distribution. The midsole is divided into five distinct areas, each corresponding to specific stress distribution patterns observed in the plantar pressure data.

Although the optimal foam density identification process thus far has only been applied to normal walking conditions for a specific subject, *i.e.*, personalized, similar optimization procedures can be applied to other biomechanical events. In short, the individualized plantar pressure distribution can be measured while considering factors such as body weight, gait speed, and terrain, resulting in a personalized midsole design to accommodate the biomechanics.

## 5. Current Limitations and Future Directions

The current study has certain limitations that must be addressed to provide a more comprehensive understanding of midsole design optimization. The present work focused only on walking and did not consider other conditions with higher applied forces, such as jogging and running (*i.e.*, involving different stress distributions, strain rates, and impact forces). Additionally, the study did not consider the shear strain and its distribution in the vicinity of density gradation, which could significantly affect the performance and durability of the midsole. From a mechanics perspective, other limitations include neglecting factors such as fatigue life (due to cyclic loading-unloading of the foam) and the influence of temperature, humidity, and other environmental factors on the mechanical and viscoelastic properties of the foam. These factors could lead to the variation of stress and strain distributions during use, potentially impacting the long-term performance and durability of the optimized midsole designs.

## 6. Conclusions

The present study demonstrated the in-plane density gradation as a practical strategy to optimize the cushioning performance of midsole foams. Polyurea foams with various densities, ranging from 95 kg/m$^3$ to 350 kg/m$^3$ were used due to their excellent mechanical load-bearing and energy absorption characteristics. Informed by *in situ* image-based measurements, realistic biomechanical loading conditions were identified and used to compare polyurea foams with conventional EVA foam counterparts in terms of mechanical energy absorption metrics. Their superior performance attributes compared to EVA foams justify the use of polyurea foams. Next, a mathematical algorithm was developed to identify the optimal polyurea foam density corresponding to specific plantar pressure zones. The optimization process used an objective function that maximized energy absorption ideality while maintaining the foam density at a minimum. In particular, it was found that specific single-density foams are optimal for different plantar zones based on local stress (pressure) levels. For example, the 95 kg/m$^3$ density foam was identified an ideal foam for stresses up to 0.039 MPa and a 140 kg/m$^3$ foam for stresses from 0.04 to 0.047 MPa, considering optimum biomechanical energy dissipation without additional weight. We incorporated local plantar peak pressure data and categorized the midsole into five areas based on distinctive stress distribution patterns. Results obtained from the optimization process indicated that the low-density (140 kg/m$^3$) polyurea foams are proper candidates for the central and lateral phalanges of the foot due to the relatively low local plantar pressures at these locations. On the other hand, the metatarsal and arch regions require higher densities in the range of 185-230 kg/m$^3$, stiffer polyurea foams capable of withstanding larger plantar pressures while providing sufficient cushioning. Although in the present paper only walking conditions were studied, the approach presented herein can be applied to various loading conditions and user/patient-specific walking and plantar pressure patterns, leading to personalized midsole designs according to the unique physical conditions of the user.

**Acknowledgments**

This research is based upon work partially supported by the National Science Foundation under grants 2035660 (Koohbor) and 2035663 (Youssef). We gratefully acknowledge the financial support provided by New Jersey Health Foundation, grant No. PC 46-22 (Koohbor & Trkov).

**Data Availability**

Data will be available upon request from the corresponding author.

**Declaration of Conflicting Interests**

The authors declare that there is no conflict of interest.

**References**

1. Roddy E, Thomas MJ, Marshall M, et al. The population prevalence of symptomatic radiographic foot osteoarthritis in community-dwelling older adults: cross-sectional findings from the clinical assessment study of the foot. Ann Rheum Dis 2015; 74: 156–163.

2. Rome K, Frecklington M, McNair P, et al. Foot pain, impairment, and disability in patients with acute gout flares: a prospective observational study. Arthritis Care Res 2012; 64: 384–388.

3. Tomassoni D, Traini E and Amenta F. Gender and age related differences in foot morphology. Maturitas 2014; 79: 421–427.

4. De Castro AP, Rebelatto JR and Aurichio TR. The relationship between foot pain, anthropometric variables and footwear among older people. Appl Ergon 2010; 41: 93–97.

5. Jellema AH, Huysmans T, Hartholt K, et al. Shoe design for older adults: evidence from a systematic review on the elements of optimal footwear. Maturitas 2019; 127: 64–81.

6. Talmo CT, Aghazadeh M and Bono JV. Perioperative complications following total joint replacement. Clin Geriatr Med 2012; 28: 471–487.

7. Waller B, Ogonowska-Slodownik A, Vitor M, et al. Effect of therapeutic aquatic exercise on symptoms and function associated with lower limb osteoarthritis: systematic review with meta-analysis. Phys Ther 2014; 94: 1383–1395.

8. Frecklington M, Dalbeth N, McNair P, et al. Footwear interventions for foot pain, function, impairment and disability for people with foot and ankle arthritis: a literature review. Elsevier 2018; 47: 814–824.

9. Wagner A, Luna S. Effect of footwear on joint pain and function in older adults with lower extremity osteoarthritis. J Geriatr Phys Ther 2018; 41: 85–101.

10. Shorten MR. The energetics of running and running shoes. J Biomech 1993; 26: 41–51.

11. Zhang JH, McPhail AJC, An WW, et al. A new footwear technology to promote non-heelstrike landing and enhance running performance: Fact or fad? J Sports Sci 2017; 35: 1533–1537.

12. Zolfagharian A, Lakhi M, Ranjbar S, et al. Custom shoe sole design and modeling toward 3D printing. Int J Bioprinting 2021; 7: 4.

13. Begg L and Burns J. A comparison of insole materials on plantar pressure and comfort in the neuroischaemic diabetic foot. Clin Biomech 2008; 23: 710–711.

14. Nigg BM, Herzog W and Read LJ. Effect of viscoelastic shoe insoles on vertical impact forces in heel-toe running. Am J Sports Med 1988; 16: 70–76.

15. Chiu HT and Shiang TY. Effects of insoles and additional shock absorption foam on the cushioning properties of sport shoes. J Appl Biomech 2007; 23: 119–127.

16. Wang L, Hong Y and Li JX. Durability of running shoes with ethylene vinyl acetate or polyurethane midsoles. J Sports Sci 2012; 30: 1787–1792.

17. Chen WM, Lee SJ and Lee PVS. Plantar pressure relief under the metatarsal heads – Therapeutic insole design using three-dimensional finite element model of the foot. J Biomech 2015; 48: 659–665.

18. Yang Z, Cui C, Wan X, et al. Design feature combinations effects of running shoe on plantar pressure during heel landing: A finite element analysis with Taguchi optimization approach. Front Bioeng Biotechnol 2022; 10.

19. Song Y, Cen X, Chen H, et al. The influence of running shoe with different carbon-fiber plate designs on internal foot mechanics: A pilot computational analysis. J Biomech 2023; 153: 111597.

20. Shimazaki Y, Nozu S and Inoue T. Shock-absorption properties of functionally graded EVA laminates for footwear design. Polym Test 2016; 54: 98–103.

21. Dong G, Tessier D and Zhao YF. Design of shoe soles using lattice structures fabricated by additive manufacturing. Cambridge Univ Press 2019; 1: 719–728.

22. Tang Y, Dong G, Xiong Y, et al. Data-driven design of customized porous lattice sole fabricated by additive manufacturing. Procedia Manuf 2021; 53: 318–326.

23. Rahman O, Uddin KZ, Muthulingam J, et al. Density-Graded Cellular Solids: Mechanics, Fabrication, and Applications. Adv Eng Mater 2022; 24: 2100646.

24. Clermont C, Barrons ZB, Esposito M, et al. The influence of midsole shear on running economy and smoothness with a 3D-printed midsole. Sports Biomech 2022; 1–12.

25. Fadeel A, Abdulhadi H, Newaz G, et al. Computational investigation of the post-yielding behavior of 3D-printed polymer lattice structures. J Comput Des Eng 2022; 9: 263–277.

26. Anni IA, Uddin KZ, Pagliocca N, et al. Out-of-plane load-bearing and mechanical energy absorption properties of flexible density-graded TPU honeycombs. Compos Part C Open Access 2022; 8: 100284.

27. Al-Ketan O and Abu Al-Rub RK. Multifunctional mechanical metamaterials based on triply periodic minimal surface lattices. Adv Eng Mater 2019; 21: 1900524.

28. Rahman O and Koohbor B. Optimization of energy absorption performance of polymer honeycombs by density gradation. Compos Part C Open Access 2020; 3: 100052.

29. Kumar A, Collini L, Daurel A, et al. Design and additive manufacturing of closed cells from supportless lattice structure. Addit Manuf 2020; 33: 101168.

30. Kolken HM, Zadpoor AA. Auxetic mechanical metamaterials. RSC Adv 2017; 7: 5111–5129.

31. Pagliocca N, Uddin KZ, Anni IA, et al. Flexible planar metamaterials with tunable Poisson's ratios. Mater Des 2022; 215: 110446.

32. Uddin KZ, Pagliocca N, Anni IA, et al. Multiscale Strain Field Characterization in Flexible Planar Auxetic Metamaterials with Rotating Squares. Adv Eng Mater 2023; 25: 2201248.

33. Amorim DJN, Nachtigall T and Alonso MB. Exploring mechanical meta-material structures through personalised shoe sole design. 2019; 1–8.

34. Ghesquière-Diérickx T, Tireford N. Exploration of auxetics materials in sport. 2021.

35. Zhan T. Progress on different topology optimization approaches and optimization for additive manufacturing: a review. IOP Publ 2021; 1939: 012101.

36. Bugin LAK, Fagundes CVM, Bruscato UM, et al. Exploration of data-driven midsole algorithm design based in biomechanics data and Voronoi 3D to digital manufacturing. Des Tecnol 2020; 10: 01–10.

37. Li D, Dai N, Jiang X, et al. Interior structural optimization based on the density-variable shape modeling of 3D printed objects. Int J Adv Manuf Technol 2016; 83: 1627–1635.

38. Wang G, Shen L, Zhao J, et al. Design and compressive behavior of controllable irregular porous scaffolds: based on voronoi-tessellation and for additive manufacturing. ACS Biomater Sci Eng 2018; 4: 719–727.

39. Wu J. Continuous optimization of adaptive quadtree structures. Comput-Aided Des 2018; 102: 72–82.

40. Echeta I, Feng X, Dutton B, et al. Review of defects in lattice structures manufactured by powder bed fusion. Int J Adv Manuf Technol 2020; 106: 2649–2668.

41. Maconachie T, Leary M, Lozanovski B, et al. SLM lattice structures: Properties, performance, applications and challenges. Mater Des 2019; 183: 108137.

42. Gupta N. A functionally graded syntactic foam material for high energy absorption under compression. Mater Lett 2007; 61: 979–982.

43. Higuchi M, Adachi T, Yokochi Y, et al. Controlling of distribution of mechanical properties in functionally-graded syntactic foams for impact energy absorption. Trans Tech Publ 2012; 706: 729–734.

44. Koohbor B and Kidane A. Design optimization of continuously and discretely graded foam materials for efficient energy absorption. Mater Des 2016; 102: 151–161.

45. Naebe M and Shirvanimoghaddam K. Functionally graded materials: A review of fabrication and properties. Appl Mater Today 2016; 5: 223–245.

46. Uddin KZ and Koohbor B. Gradient optimization of transversely graded Ti-TiB structures for enhanced fracture resistance. Int J Mech Sci 2020; 187: 105917.

47. Uddin KZ, Anni IA, Youssef G, et al. Tuning the Mechanical Behavior of Density-Graded Elastomeric Foam Structures via Interlayer Properties. ACS Omega 2022; 7: 37189–37200.

48. Smeets M, Koohbor B and Youssef G. Quasi-Static Mechanical Response of Density-Graded Polyurea Elastomeric Foams. ACS Appl Polym Mater 2023.

49. Gupta V and Youssef G. Orientation-dependent impact behavior of polymer/EVA bilayer specimens at long wavelengths. Exp Mech 2014; 54: 1133–1137.

50. Tang L, Wang L, Bao W, et al. Functional gradient structural design of customized diabetic insoles. J Mech Behav Biomed Mater 2019; 94: 279–287.

51. Mahesh CC and Ramachandran KI. Finite element modelling of functionally graded elastomers for the application of diabetic footwear. Mater Today Proc 2018; 5: 16367–16377.

52. Worobets J, Wannop JW, Tomaras E, et al. Softer and more resilient running shoe cushioning properties enhance running economy. Footwear Sci 2014; 6: 147–153.

53. Sterzing T, Custoza G, Ding R, et al. Segmented midsole hardness in the midfoot to forefoot region of running shoes alters subjective perception and biomechanics during heel-toe running revealing potential to enhance footwear. Footwear Sci 2015; 7: 63–79.

54. Ali M, Nazir A and Jeng JY. Mechanical performance of additive manufactured shoe midsole designed using variable-dimension helical springs. Int J Adv Manuf Technol 2020; 111: 3273–3292.

55. Uddin KZ, Youssef G, Trkov M, et al. Gradient optimization of multi-layered density-graded foam laminates for footwear material design. J Biomech 2020; 109: 109950.

56. Brückner K, Odenwald S, Schwanitz S, et al. Polyurethane-foam midsoles in running shoes-impact energy and damping. Procedia Eng 2010; 2: 2789–2793.

57. Verdejo R, Mills N. Performance of EVA foam in running shoes. Eng Sport 2002; 4: 580–587.

58. Prajapati MJ, Kumar A, Lin SC, et al. Multi-material additive manufacturing with lightweight closed-cell foam-filled lattice structures for enhanced mechanical and functional properties. Addit Manuf 2022; 54: 102766.

59. Reed N, Huynh NU, Rosenow B, et al. Synthesis and characterization of elastomeric polyurea foam. J Appl Polym Sci 2020; 137: 48839.

60. Youssef G, Kokash Y, Uddin KZ, et al. Density-Dependent Impact Resilience and Auxeticity of Elastomeric Polyurea Foams. Adv Eng Mater 2023; 25: 2200578.

61. Koohbor B, Blourchian A, Uddin KZ, et al. Characterization of energy absorption and strain rate sensitivity of a novel elastomeric polyurea foam. Adv Eng Mater 2021; 23: 2000797.

62. Youssef G and Reed N. Scalable manufacturing method of property-tailorable polyurea foam. Google Patents, 2021.

63. Do S, Huynh NU, Reed N, et al. Partially-perforated self-reinforced polyurea foams. Appl Sci 2020; 10: 5869.

64. Sutton MA, Orteu JJ and Schreier H. Image correlation for shape, motion and deformation measurements: basic concepts, theory and applications. Springer Sci Bus Media 2009.

65. Youssef G. Applied mechanics of polymers: properties, processing, and behavior. Elsevier 2021.

66. Koohbor B, Kidane A and Lu WY. Characterizing the constitutive response and energy absorption of rigid polymeric foams subjected to intermediate-velocity impact. Polym Test 2016; 54: 48–58.

67. Whitten I and Youssef G. The effect of ultraviolet radiation on ultrasonic properties of polyurea. Polym Degrad Stab 2016; 123: 88–93.

68. Youssef G and Whitten I. Dynamic properties of ultraviolet-exposed polyurea. Mech Time-Depend Mater 2017; 21: 351–363.

69. Koohbor B, Youssef G, Uddin KZ, et al. Dynamic behavior and impact tolerance of elastomeric foams subjected to multiple impact conditions. J Dyn Behav Mater 2022; 8: 359–370.